%Paper: adap-org/9409002
%From: juriev@physique.ens.fr (Denis Juriev)
%Date: Thu, 8 Sep 1994 11:01:48 --100
%Date (revised): Fri, 9 Sep 1994 09:21:17 --100
%Date (revised): Fri, 9 Sep 1994 09:51:10 --100
%Date (revised): Thu, 22 Sep 94 16:05:44 +0200

\input amstex
\magnification=1200
\documentstyle{amsppt}
\font\cyr=wncyr10
\font\cyre=wncyr8
\font\cyb=wncyb10
\NoRunningHeads
\NoBlackBoxes
\define\SU{\operatorname{\bold S\bold U}}
\topmatter
\title\cyb
Vtorichnyi0 sintez izobrazhenii0 v e1lektronnoi0 kompp1yuternoi0 fotografii.
\endtitle
\author\cyr Yurp1ev D.V.
\endauthor
\address\cyre\newline Otdel matematiki,\newline
Nauchno-issledovatelp1skie0 institut\newline
sistemnyh issledovanie0 RAN,\newline
Moskva
\endaddress
\date adap-org/9409002\enddate
\endtopmatter
\document\cyr

Osnovnaya ideya vtorichnogo sinteza izobrazhenie0 v e1lektronnoe0
kompp1\-yu\-ternoe0 fotografii sostoit v preobrazovanii raznorodnyh, no prosto
organizovannyh vhodnyh vizualp1nyh dannyh (poluchennyh v razlichnye momenty
vremeni v razlichnyh rakursah i podvergnutyh "standartnomu" pervichnomu
sintezu) v odnorodnye vyhodnye dannye bolee slozhnoe0 struktury v
nestandartnoe0 tsvetoperspektivoe0 sisteme [1]. Ispolp1zovanie nestandartnyh
tsvetoperspektivnyh sistem i, v chastnosti, anomalp1nyh tsvetovyh prostranstv,
soderzhashchih naryadu s tremya osnovnymi bazisnymi tsvetami dostatochnoe
kolichestvo obertsvetov pozvolyaet polnostp1yu so\-hra\-nitp1 vhodnuyu
informatsiyu pri ukazannom preobrazovanii. Printsi\-pi\-alp1naya s\-hema
e1lektronnoe0 kompp1yuternoe0 fotografii, osnovannoe0 na vto\-rich\-nom sinteze
izobrazhenie0, imeet vid:
$$\align\boxed{\aligned &\text{\cyr Videokompp1yuter}\\
&\text{\rm (Videocomputer)}\endaligned}&\longleftarrow
\boxed{\aligned &\text{\cyr Kompp1yuternye0 anomalae0zer}\\
&\text{\rm (computer anomalizer)}\endaligned}\longleftarrow\\
&\longleftarrow\boxed{\aligned &\text{\cyr Detektory}\\
&\text{\cyr [prinimayushchie ustroe0stva, fotokamery]}\\
&\text{\rm (detectors)}\endaligned}\endalign$$

Ispolp1zovanie neskolp1kih detektorov (prinimayushchih ustroe0stv)
poz\-vo\-lya\-et
uchestp1 stereoe1ffekty, v to vremya kak sp2i0mka v razlichnye momenty vremeni
--- dvizhenie snimaemogo obp2ekta. Kompp1yuternye anomalae0zery ({\rm computer
anomalizers}) preobrazuyut po opredeli0nnomu pravilu vhodnye dannye,
predstavlennye v tsifrovoe0 forme, v vyhodnye dannye, vo\-spro\-iz\-vo\-di\-mye
na
e1krane videokompp1yutera. Vhodnaya cifrovaya informatsiya mozhet
podvergatp1sya vtorichnomu sintezu kak v moment sp2i0mki, tak i v moment
vosproizvedeniya. V zavisimosti ot vybora s\-hemy kompp1yuternye0 anomalae0zer
yavlyaet\-sya libo pristavkoe0 k tsifrovoe0 fotokamere, libo so\-stav\-noe0
chastp1yu programmnogo obespecheniya videokompp1yutera.

\define\inp{\operatorname{input}}
\define\outp{\operatorname{output}}
Razberi0m process vtorichnogo sinteza na primere tsvetoperspektivnoe0 sistemy
"podvizhnogo videniya" ({\rm mobilevision, MV}) [1]. Vhodnye dannye v e2tom
sluchae
opisyvayut\-sya linee0nym prostranstvom $V_{\inp}=\oplus_{a\in A}V_a$, gde
$V_a$
-- linee0nye prostranstva, izomorfnye obychnomu tsvetovomu prostranstvu $V$,
indeks $a$ harakterizuet prinimayushchee ustroe0stvo i moment sp2i0mki.
Vyhodnye dannye opisyvayut\-sya anomalp1nym tsvetovym pro\-stran\-s\-t\-vom
$V_{\outp}$,
predstavlyayushchim soboe0 proektivnye0 $\SU(3)$--gipermulp1tip\-let. takim
obrazom, kompp1yuternye0 anomalae0zer osushchestvlyaet linee0noe
preobrazovanie $V_{\inp}$ v $V_{\outp}$. Dlya togo chtoby nae0ti dopustimoe
mno\-zhe\-s\-t\-vo linee0nyh operatorov $D$ iz $V_{\inp}$ v $V_{\outp}$,
osushchestvlyayushchih vtorichnye0 sintez izobrazheniya, neobhodimo uchestp1
vnutrennie simmetrii $V_{\inp}$ i $V_{\outp}$. Estestvenno predpolagatp1, chto
$D$ yavlyaet\-sya $\SU(3)$--spletayushchim operatorom [2]; tem samym operator
$D$ stroit\-sya na baze koe1fficientov Klebsha--Gordana gruppy $\SU(3)$ [3].

Rassmotrim sluchae0 $^t\SU(3)$--{\rm WZNW}--tsvetovogo prostranstva [1] v
ka\-che\-s\-t\-ve
primera. V e1tom sluchae, kak pravilo, opredelena estestvennaya proektsiya
$D_0:T(V)\mapsto V_{\outp}$, gde $T(V)$ -- summa tenzornyh stepenee0
prostranstva $V$. Operator $D_0$ neposredstvenno vyrazhaet\-sya cherez
koe1ffi\-tsi\-en\-ty Klebsha--Gordana gruppy $\SU(3)$, v to vremya kak operator
$D$
oprede\-lya\-et\-sya polinomom $P(x_1,\ldots x_N)$ ot nekommutiruyushchih
peremennyh
$x_a$, otve\-cha\-yu\-shchih vhodnym prostranstvam $V_a$, a imenno, esli $v_a$
---
sovokupnostp1 e1lementov iz $V_a$, to $D(v_1,\ldots v_N)=D_0(P(v_1,\ldots
v_N))$.

Dlya togo chtoby nae0ti dopustimoe mnozhestvo mnogochlenov $P$ i
ope\-ra\-to\-rov
$D$ neobhodimo uchestp1 strukturu vhodnyh dannyh,
harak\-terizu\-yu\-shchu\-yu\-sya
prostranstvennymi, vremennymi i skrytymi simmetriyami.

Prostranstvennye simmetrii opredelyayut\-sya konfiguratsiee0
prinima\-yu\-shchih
ustroe0stv. Naprimer, v binokulyarnom sluchae imeet\-sya $\Bbb
Z_2$--simmet\-ri\-ya,
a v die1dralp1nom geksagonalp1nom sluchae $D_6$--simmetriya, takim obrazom v
naibolee interesnyh sluchayah geometricheskie konfiguracii detektorov
opisyvayut\-sya konechnymi gruppami [4].

Vremennye simmetrii mogut bytp1 rassmotreny analogichno.

Esli simmetrii obrazuyut gruppu $G$, to e1ta gruppa obyazana dee0stvo\-vatp1 v
proektivnom $\SU(3)$--gipermulp1tiplete $V_{\outp}$ avtomorfizmami.\linebreak
Ukazannoe
trebovanie suzhaet krug dopustimyh anomalp1nyh tsvetovyh\linebreak prostranstv,
a takzhe
nakladyvaet na operator $D$ dopolnitelp1noe uslo\-vie: on dolzhen bytp1 ne
tolp1ko $\SU(3)$--spletayushchim, no i $G$--spletayushchim operatorom; kak
sledstvie, pri postroenii mnogochlena $P$ ispolp1zuyut\-sya
koe1ffitsienty Klebsha--Gordana konechnoe0 gruppy $G$.

Issledovanie vozmozhnyh skrytyh simmetrie0 predstavlyaet soboe0 interesnuyu
zadachu.
\newpage

\Refs
\roster
\item"[1]" {\rm Juriev D., Anomalous color spaces and their structure. The
Visual Computer, 1994 (to appear);\newline Octonions and binocular
Mobilevision;
{\it hep-th/9401047}.
\item"[2]" Lenz R., Group theoretical methods in image processing, Springer,
1990;\newline Kanatani K.I., Group theoretical methods in image understanding,
Springer, 1991.}
\item"[3]" {\cyre Klimyk A.U., Matrichnye e1lementy i koe1ffitsienty
Klebsha--Gordana predstavlenie0 grupp. Kiev, Naukova Dumka, 1979.
\item"[4]" Kovali0v O.V., Neprivodimye i indutsirovannye predstavleniya i
ko\-pred\-stav\-le\-ni\-ya fi0dorovskih grupp. M., Nauka, 1986.}
\endroster
\endRefs
\enddocument